\documentclass{article} 
\usepackage{icomp2024_conference,times}

\usepackage{amsmath,amsfonts,bm}

\def\eqref#1{equation~\ref{#1}}

\def\1{\bm{1}}

\DeclareMathAlphabet{\mathsfit}{\encodingdefault}{\sfdefault}{m}{sl}
\SetMathAlphabet{\mathsfit}{bold}{\encodingdefault}{\sfdefault}{bx}{n}

\usepackage{makecell}
\usepackage{pgfplots}
\usepackage{pgfplotstable}
\usepackage{filecontents}
\usepackage{hyperref}
\usepackage{url}
\usepackage{pgfplots}
\usepackage{appendix}
\usepackage{afterpage}

\AtBeginDocument{%
  \providecommand\BibTeX{{%
    \normalfont B\kern-0.5em{\scshape i\kern-0.25em b}\kern-0.8em\TeX}}}

\title{Finetuning Large Language Models for Vulnerability Detection}

\icompfinalcopy

\author{
Aleksei Shestov, Rodion Levichev, Ravil Mussabayev, Evgeny Maslov, Anton Cheshkov \& Pavel Zadorozhny \\
\And
Aleksei Shestov \thanks{The work was done while the author was at the Huawei Russian Research Institute} \\
Sber AI Lab \\
Moscow 117997, Russia \\
\texttt{AMShestov@sberbank.ru,shestovmsu@gmail.com} \\
\And
Rodion Levichev \\
Huawei Russian Research Institute \\
Moscow 121099, Russia \\
\texttt{rodion.levichev@huawei-partners.com} \\
\And
Ravil Mussabayev \thanks{The work was done while the author was at the Huawei Russian Research Institute} \\
Satbayev University \\
Almaty 121099, Kazakhstan \\
\texttt{r.mussabayev@satbayev.university} \\
\And
Evgeny Maslov \\
Huawei Russian Research Institute \\
Moscow 121099, Russia \\
\texttt{evgeny.maslov@huawei.com} \\
\And
Anton Cheshkov \\
Huawei Russian Research Institute \\
Moscow 121099, Russia \\
\texttt{anton@huawei.com} \\
\And
Pavel Zadorozhny \\
Huawei Russian Research Institute \\
Moscow 121099, Russia \\
\texttt{pavel.zadorozhny@huawei-partners.com}
}

\begin{document}

\maketitle

\begin{abstract}
This paper presents the results of finetuning large language models (LLMs) for the task of detecting vulnerabilities in Java source code. We leverage WizardCoder, a recent improvement of the state-of-the-art LLM StarCoder, and adapt it for vulnerability detection through further finetuning. To accelerate training, we modify WizardCoder's training procedure, also we investigate optimal training regimes. For the imbalanced dataset with many more negative examples than positive, we also explore different techniques to improve classification performance. The finetuned WizardCoder model achieves improvement in ROC AUC and F1 measures on balanced and imbalanced vulnerability datasets over CodeBERT-like model, demonstrating the effectiveness of adapting pretrained LLMs for vulnerability detection in source code. The key contributions are finetuning the state-of-the-art code LLM, WizardCoder, increasing its training speed without the performance harm, optimizing the training procedure and regimes, handling class imbalance, and improving performance on difficult vulnerability detection datasets. This demonstrates the potential for transfer learning by finetuning large pretrained language models for specialized source code analysis tasks.
\end{abstract}

\section{Introduction}

Detecting vulnerabilities in source code is an important problem in software engineering. This project explores finetuning of LLMs for binary classification of Java functions as vulnerable or not. The goals are the following:
\begin{itemize}
    \item Investigate the possibility of applying the new class of big LLM models for the task of vulnerability detection;
    \item Improve over previous results of CodeBERT-style models by using big LLM models;
    \item Investigate whether the encountered performance limit is due to the limited capacity of CodeBERT-like models.
\end{itemize}


Transformer models pretrained on large code corpus, like  CodeBERT \cite{Feng2020-codebert} and ContraBERT \cite{Liu2023-contrabert}, have shown promising results on many tasks of code understanding. LineVul \cite{Fu2022-linevul}, a model built on CodeBERT, achieved state-of-the-art results on the vulnerability detection task, surpassing graph network approaches, like Reveal \cite{Chakraborty2022-reveal}. Recent LLMs, which have billions of parameters, such as StarCoder \cite{Li2023-starcoder} and WizardCoder \cite{Luo2023-wizardcoder}, are pretrained on trillions of tokens. They are 2 orders of magnitude bigger than CodeBERT (CodeBERT has approximately 125 million parameters) and have orders of magnitude more data used for pretraining. They achieve better results on language and code understanding tasks, providing an opportunity for improved transfer learning. The vulnerability detection task can benefit from the increased expressiveness and knowledge contained in state-of-the-art LLMs.

Vulnerability detection is a complex task with many aspects influencing and potentially limiting the solution quality. The problem of function-level vulnerability detection can be divided into two stages: representing the function in some form (such as execution paths or various graphs) and solving the classification problem for this representation~\cite{Chakraborty2022-reveal,Zhang2023-epvd}. It is noted that each stage can be independently improved, and improving each stage enhances the overall quality of the solution to the problem. Justifications to these observations include: the EPVD model~\cite{Zhang2023-epvd}, where improving the data structure with the same ML algorithm resulted in gains, and the LineVul model~\cite{Fu2022-linevul} compared to articles that present the function text to an LSTM. Therefore, it is relevant to improve each of the two stages.

We focus on improving the second part, the ML algorithm for vulnerability detection, without addressing the first stage. Our approach involves two directions: enhancing the model's capacity and improving the handling of imbalanced data, which represents real-world conditions. One of the limiting factors could be the capacity and code understanding ability of CodeBERT-like models, as vulnerability detection requires a deep understanding of many code aspects. Checking this hypothesis is a crucial problem in the field of vulnerability detection. Hence, we compare our classification of functions, represented as text, with the previous most powerful family of models, the CodeBERT-based models.


Furthermore, the study by \cite{Cheshkov2023-chatgptvuleval} revealed that simple prompting techiques for the ChatGPT LLM were unable to achieve noticeably better results than random prediction on the vulnerability detection task. This finding suggests that finetuning of LLMs on this task naturally becomes a promising research direction.


The real-world distribution of vulnerable functions is highly class imbalanced, containing far more negative samples than positive ones. This imbalance makes the task inherently more difficult, as standard training objectives, like cross-entropy loss, may lead the model to focus on the dominant negative class. Specialized training techniques, such as focal loss \cite{Lin2017-focalloss} and sample weighting, are required to properly emphasize the minority positive examples during finetuning.

This project conducts experiments to find the optimal model architecture, training hyperparameters, and loss formulations for effective LLM finetuning on the vulnerability detection task (both for balanced and imbalanced datasets). The aim is to leverage recent advances in pretrained models to improve over prior benchmark results. 

The remainder of the paper is organized as follows:
\begin{itemize}
\item We discuss the most notable related work in the field;
\item We describe our approach for building and adapting the model. Specifically, we discuss how we solved the following problems:
\begin{itemize}
\item Choosing the model;
\item Using limited GPU resources for the model size;
\item Adapting a decoder model for the classification task;
\item Underutilization of the input sequence length, resulting in a slow training speed.
\end{itemize}
\item We detail our main instruments: datasets, models, metrics, as well as measurement procedures.
\item We describe the conducted experiments:
\begin{itemize}
\item Comparison with ContraBERT (state-of-the-art non-LLM transformer model) on balanced and imbalanced datasets;
\item Ablation study of the context size and classification loss;
\item Investigation of the batch packing technique;
\item Improving the usage of a more informative part of the imbalanced dataset.
\end{itemize}
\item We conclude the study, highlighting our contributions and stating promising future research directions.
\end{itemize}

Appendix~\ref{sec:cwe_distribution} provides additional details about the obtained distribution of CWE types in our dataset.

\section{Related work}

Recent work by Zhou et al. \cite{Zhou2024LargeLM} provides a comprehensive review of Large Language Models (LLMs) for vulnerability detection and repair. Their analysis reveals rapid growth in this field, with 27.8\% of reviewed studies published in just the first two months of 2024. The majority of approaches (82\%) employ fine-tuning techniques, while zero-shot (11\%) and few-shot (7\%) methods are less common. CodeBERT emerges as the dominant model for vulnerability detection, used in 38.5\% of studies, with GPT-3.5 and GPT-4 following at 10.3\% and 2.6\% respectively. The review also highlights persistent challenges in the field, including class imbalance, noisy data, and scarcity of labeled datasets for training LLMs in vulnerability detection. Our work addresses these challenges by focusing on fine-tuning WizardCoder for Java vulnerability detection, contributing to this rapidly evolving research area.

Despite the successes of GPT models in solving various code-related tasks, their ability to detect vulnerabilities in code remains insufficient. The report "Evaluation of ChatGPT Model for Vulnerability Detection" \citep{Cheshkov2023-chatgptvuleval} notes that ChatGPT and GPT-3 models, without additional tuning and training, show results no better than random guessing in vulnerability classification. Similar conclusions are drawn in the study "ChatGPT for Vulnerability Detection, Classification, and Repair: How Far Are We?" (Michael Fu et al.) \citep{10479409}, where comparisons with specialized models such as CodeBERT and GraphCodeBERT reveal a significant lag of ChatGPT in vulnerability-related tasks. 

Another notable example is the study by Liu et al. (2023) \citep{10381286}, where the GPT model was enhanced using retrieval-augmented generation (RAG). Techniques such as BM-25 and TF-IDF were employed to retrieve relevant information, which was then used as examples for the few-shot learning technique. Using the Devign dataset, the study demonstrated significant improvements in precision, recall, and F1 score compared to traditional methods. 

Recent work by Purba et al. \cite{10301302} explores the application of large language models (LLMs) for software vulnerability detection in C and C++ code. They evaluate four LLMs, including fine-tuned and zero-shot models, on SQL injection and buffer overflow vulnerabilities using the Code Gadgets and CVEfixes datasets. Their findings reveal that while LLMs demonstrate high recall in identifying vulnerable code patterns, they suffer from high false positive rates. This aligns with our observations on the challenges of using LLMs for vulnerability detection. However, our study extends this approach to Java, a domain where LLM-based vulnerability detection remains underexplored. We focus specifically on fine-tuning the state-of-the-art WizardCoder model for Java code analysis and introduce novel techniques to improve training efficiency and handle class imbalance. Unlike Purba et al., who primarily use prompt-based approaches, we explore more advanced fine-tuning methods and provide a detailed analysis of the model's performance on Java-specific vulnerability types, contributing to the limited body of work on LLM-based vulnerability detection for Java.

\section{Implementation details} \label{appx:impl_details}

\subsection{Challenges}

We briefly describe the main milestones and challenges that we encountered during the implementation of our approach.

\paragraph{LLM model selection for finetuning}

Recently, there has been a rise in the number of open-sourced LLM models, trained on large code corpora \cite{Li2023-starcoder, Luo2023-wizardcoder, Zheng2023-codegeex, Nijkamp2023-codegen2}. The task is to choose an optimal model for finetuning, both from the perspective of quality and potential technical difficulties encountered. The goal is to select a model possessing the following properties:

\begin{itemize}
\item It shows a strong potential for effective transfer learning to the vulnerability detection task;
\item It is technically easy to use and integrate into our framework.
\end{itemize}

\paragraph{Adapting LLMs for limited computational resources}

Given our computing capabilities, finetuning complete LLM models is infeasible and would be inefficient even with increased resources. To enable effective finetuning of a 13-billion parameter model on our hardware, we need methods to reduce the number of trained parameters and memory requirements.

\paragraph{Adapting LLMs for the classification task}

The standard pretraining task for LLM decoders is the next token prediction, which differs from our end task of vulnerability classification. Including the next token prediction component in the loss may misguide weight updates away from the optimal state for classification. To fully leverage the capabilities of the LLM weights for our target task, a better strategy might be to leave only the classification term in the loss.

\paragraph{Improving a slow training speed}

The standard sequence classification approach of padding to a fixed length makes training extremely slow. To mitigate the issue of short sequence lengths, the task is to pack multiple functions into each training sequence. This can decrease unused padded tokens, improving computational efficiency.

\subsection{Choosing an LLM to finetune}

Recently, there has been a proliferation of open-sourced LLM models trained on large code corpora \cite{Li2023-starcoder, Luo2023-wizardcoder, Zheng2023-codegeex, Nijkamp2023-codegen2}. The task is to select an optimal model for finetuning.

First, we attempted to use the CodeGeeX model \cite{Zheng2023-codegeex}. Unfortunately, we found that using this model was very difficult from the technical perspective, as the code provided by the authors does not support standard libraries for transformer training and large model adaptation. We were able to run the model on 8 V100 GPUs for inference, but this memory capacity was insufficient for finetuning. The lack of support for common LLM adaptation libraries made it very challenging to overcome the memory limitations, so we decided to discontinue attempts with this model.

Next, we explored two other models, WizardCoder \cite{Luo2023-wizardcoder} and CodeGen \cite{Nijkamp2023-codegen2}. In contrast to CodeGeeX, these models include support for major transformer training frameworks like DeepSpeed, HuggingFace Transformers, Peft \cite{peft}, etc. This compatibility with standard libraries significantly eased the application of low-memory adaptation techniques.

When conducting a simple Question Answering training, we find that WizardCoder is much better than CodeGen at directly answering ``YES'' and ``NO''. CodeGen sometimes responded with ``TRUE'', ``FALSE'', erratic ``YES'' and ``NO'' with additional characters, and has a worse sequence stopping behavior. Given the WizardCoder's stronger capabilities on this task, we focused our subsequent experiments on this model.

\paragraph{Choosing the best StarCoder-family model}

In our study, we evaluate three models from the StarCoder family for finetuning: StarCoderBase, StarCoder \cite{Li2023-starcoder} (which is StarCoderBase finetuned on the Python dataset), and WizardCoder \cite{Luo2023-wizardcoder} (which is an improved version of StarCoder trained using the eval-instruct technique). We compared the performance of StarCoder and StarCoderBase on the question answering task and found no significant differences between the two models, despite the fact that StarCoder was specifically adapted for Python. Subsequently, we primarily used the WizardCoder model, as it is claimed to be superior to StarCoder. However, our experiments did not reveal any significant differences in performance between these two models. In our future writing, we will use the term ``StarCoder'' to refer specifically to the WizardCoder variant of the StarCoder family.

Generally, more elaborate experiments are needed to find if there are any differences in the performance of these models.

\subsection{Adapting LLMs for limited computational resources}

Finetuning complete multi-billion parameter LLMs is infeasible given our hardware constraints, even if more resources were obtained. To enable effective finetuning, we require techniques to reduce the number of trained parameters and memory needs. The HuggingFace Peft \cite{peft} library implements several methods tackling this, including the LORA method \cite{Hu2022-lora}. LORA enables full-model finetuning by decomposing the weight matrices into low-rank approximations, drastically decreasing parameters and memory.

We successfully applied LORA to the WizardCoder model. We used the following optimal LORA settings: r = 8, alpha = 32, dropout = 0.05. This reduced the 13 billion parameter models to just 25 million parameters, smaller than CodeBERT. The results validate LORA's effectiveness for LLM adaptation under memory limitations.


\subsection{Adapting LLMs for classification}


Typically, code LLMs are trained on a large amount of unlabeled code data in a special way. The principle is to iteratively take code tokens as input, predict the next token, and compare it with the ground truth. This principle drives the next token prediction loss. Specifically, for any input sequence $\{x_1, x_2, \ldots, x_N\}$ of length $N$, the output of an LLM is a probability distribution of the next token $P(x_{N+1} | x_1, x_2, \ldots, x_N, \Theta) = p_{N+1} \in [0, 1]^v$, where $\Theta$ represents all parameters of the model, and $v$ is the vocabulary size. By comparing it with the real distribution, i.e., a one-hot vector $y_{N+1} \in \{0, 1\}^v$ of the ground-truth token, we can optimize the cumulative cross-entropy loss:

\begin{equation}
\mathcal{L}_{\text{ntp}} = -\sum_{n=1}^{N-1} y_{n+1} \log \mathbb{P}(x_{n+1} | x_1, x_2, \ldots, x_n, \Theta) \label{eq:ntp_loss}
\end{equation}

In the context of adapting a language model for classification tasks, the objective function used during training needs to be aligned with the classification goal rather than the next token prediction. For an input sequence $\{x_1, x_2, \ldots, x_N\}$ of length $N$, the traditional generative pretraining objective (next token prediction loss) would compute the loss across all predicted tokens in the sequence. However, this is suboptimal for a classification task such as vulnerability classification, which is concerned with categorizing the entire input sequence rather than predicting subsequent tokens.

To address this, we propose a classification loss that leverages only the predicted probability of the final token $x_N$, which is then matched against the ground truth label $y$ using cross-entropy. This loss is expressed mathematically as:

\begin{equation}
\mathcal{L}_{\text{class}} = -\log \mathbb{P}(y | x_1, x_2, \ldots, x_N, \Theta) \label{eq:class_loss}
\end{equation}

Here, $y$ represents the correct class label for the sequence, and $\mathbb{P}(y | x_1, x_2, \ldots, x_N, \Theta)$ is the probability that the model, parameterized by $\Theta$, assigns to the correct class. This formulation ensures that weight updates during training are driven exclusively by the classification objective, without any influence from the generative task of next token prediction. The classification loss may be viewed as corresponding to the last term of the generative pretraining objective's (next token prediction) cross-entropy loss, focusing entirely on the classification output for $n = N$.

Eliminating the next token prediction objective enables full utilization of the pretrained weights for vulnerability detection, without interference from unrelated generation tasks.



\subsection{Speeding up the classification training}

The standard approach of passing code samples to the LLM by writing tokens into the context window and padding unused slots is problematic for our data. Most dataset functions have length under 50 tokens (see Table~\ref{fig:func_len_hist}), so this padding wastes computation.

\begin{table}[ht]
\centering
\begin{tabular}{ll}
\hline
Function length (tokens) & Functions number \\
\hline
0-10   & 1976 \\
10-20 & 31627 \\
20-50 & 51972 \\

50-100 & 32194 \\
100-300 & 35769 \\
300-500 & 8756 \\
500-1000 & 5515 \\

1000-2000 & 2236 \\
$>$2000 & 767 \\
\hline
\end{tabular}
\caption{Function length histogram, obtained from parsing several projects}
\label{fig:func_len_hist}
\end{table}

We see that most functions have less than 50 tokens, with many having 10-20 tokens. Using one function per batch element (padded to 2048) is highly redundant.

To mitigate the issue of short sequence lengths, we pack multiple functions into each training sequence. This increases the actual batch size versus using one function per sequence.

Thus, during our study we focus on the following two regimes of using code LLMs:
\begin{enumerate}
    \item \textit{Next token prediction.} This approach corresponds to using the next token prediction loss~\eqref{eq:ntp_loss} during training and prediction. During training, an LLM's input sequence is packed with the code of training methods and their ground truth labels one after another in the following format:
    
    \vspace{5pt}
    Method's code + YES / NO (binary ground truth label) + $\langle \text{EOS} \rangle$
    \vspace{5pt}

    Here $\langle \text{EOS} \rangle$ denotes the special ``end of sequence'' token from the vocabulary. We fit as many as possible full training examples into the input sequence length, which is equal to $2048$ for StarCoder-based models and $512$ for CodeBERT-based ones. The rest of the positions are filled with the special padding token. For prediction, the binary classification token is generated according to \eqref{eq:ntp_loss}, and its probability is taken for ROC AUC calculation;

    \item \textit{Binary classification.} This regime uses the binary cross entropy classification loss \eqref{eq:class_loss} for training and prediction. Specifically, the training scheme for binary classification loss is formalized as follows:

    \begin{enumerate}
        \item Load a batch with as many complete function codes as it can fit in the following format:

        \vspace{5pt}
        Method's code + $\langle \text{EOS} \rangle$
        \vspace{5pt}
        
        \item Populate an array of labels, assigning a label to each function in the batch, where the label is either $1$ or $0$, indicating whether the function is vulnerable or not, respectively;
        
        \item During loss computation, match the predicted probability of the next token for the last token of each function with its corresponding label from the label array;
        
        \item Calculate the cross-entropy loss for each pair of predicted probability and actual label using \eqref{eq:class_loss};
        
        \item Sum up the cross-entropy loss values across all predictions in the batch (\textbf{not} the average, which usually works worse, as will be investigated later);
        
        \item The resulting sum is the loss for the batch, which is used to update the model's weights during training.
    \end{enumerate}
    
    The loss $\mathcal{L}_{\text{batch}}$ for a batch is thus given by:
    
    \begin{equation}
    \mathcal{L}_{\text{batch}} = \sum_{i=1}^{B} -y_i \cdot \log(p_i) - (1 - y_i) \cdot \log(1 - p_i)
    \end{equation}
    
    where $B$ is the number of functions in the batch, $y_i$ is the true label for the $i$-th function, and $p_i$ is the model's predicted probability that the $i$-th function is vulnerable.

    Due to the memory limits of our hardware, we limited the batch size to only one input sequence during training.

    For prediction, no batch packing is used and only one test function is fitted into the entire input sequence, with the rest of the token positions filled by the padding token.

\end{enumerate}

The baseline classification approach achieved only 1500 training samples per hour. This extremely slow training throughput made iterative experiments and tuning infeasible on our dataset. In contrast, batch packing provided a 13x speedup to 20000 samples per hour. By concatenating multiple short sequences into one input example, we can significantly boost efficiency for tasks like ours with small input functions. The key advantage of batch packing is reducing unused padding tokens. By packing sequences, a much larger proportion of computation goes towards informative function tokens rather than wasted padding.

In our case, batch packing enables practical finetuning by speeding up training by an order of magnitude. This allows reasonable iteration times for experimentation. Also, this approach is generally applicable when finetuning decoder models on tasks with short input sequences.

There is a room for further enhancement of the batch packing method. The dynamic batch size could be stabilized using dynamic gradient accumulation steps. The dynamic accumulation would accumulate gradients until a target batch size is reached before applying an optimizer step. This would stabilize batch size while retaining the efficiency benefits of packing sequences.

However, this poses some technical challenges. It requires non-trivial modifications to the training loop code in HuggingFace Transformers to support dynamic gradient accumulation.

\section{Experimental setup} \label{appx:exp_setup}

\subsection{Datasets}

\paragraph{Sources}

Our vulnerability dataset is constructed from several open-source vulnerability datasets: CVEfixes \cite{Bhandari2021-cvefixes}, a Manually-Curated Dataset \cite{Ponta2019-mancurdataset} and VCMatch \cite{Wang2022-vcmatch}.

CVEfixes dataset includes information about vulnerabilities and their fixes in open-source software, with data collected from the NVD database and other repositories. However, some inconsistencies and errors may be present due to the automatic construction process.

A Manually-Curated Dataset focuses on Java language only and includes data for 624 publicly disclosed vulnerabilities across 205 distinct open-source Java projects. The dataset is considered more reliable as fix-commits are found by experts.

VCMatch provides a ranking-based approach for automatic security patches localization for OSS vulnerabilities. It includes data for only 10 popular repositories and only contains fix-commits.

\paragraph{Dataset labeling}

Each of the aforementioned datasets have standalone functions as their elements. In each of the datasets functions are labeled as vulnerable or non-vulnerable based on heuristics derived from analyzing vulnerability-fixing commits:

\begin{enumerate}
\item From vulnerability-fixing commits, the changed functions are extracted;
\item The pre-change versions are taken as vulnerable functions, while the post-change versions are taken as non-vulnerable ones;
\item Pairs of changed functions are used to create a balanced dataset. Functions that remained unchanged are labeled as non-vulnerable as well.
\end{enumerate}

\paragraph{Dataset filtering}

Vulnerability-fixing commits often contain changes related to cleanups, refactors, irrelevant functionality changes. Therefore, the described labeling heuristics produce some amount of false positive labels. To address this issue, we create a modified dataset: we extract a function as vulnerable from a vulnerable-fixing commit if this function is the only one that has changed in the commit. We call this dataset $X_1$. This dataset have the following advantages:

\begin{itemize}
    \item Its labels are more robust, as commits fixing only one function are more likely to fix only vulnerabilities;
    \item The dataset is small, which makes it a good instrument for investigating the algorithmic performance and finding optimal parameters.
\end{itemize}

More specifically, extracting functions from the commits fixing only one function mitigates labeling errors that stem from irrelevant code and cleanup changes \cite{Croft2023-dataquality}.

\paragraph{Addition of easy negative functions}

As a result of these procedures, we obtain a set of vulnerable functions (we call it $P_1$) and a set of vulnerable functions with fixed vulnerability (we call it $P_2$). 
We augmented these sets with a set of unchanged functions scratched from files associated with patch, which do not have any vulnerabilities. We call this part $P_3$. 

\paragraph{Datasets characteristics}

Finally, we obtain the following two datasets:

\begin{itemize}
    \item $X_1$ without $P_3$.
    
    It has 1334 samples, with 810 samples in the train part, 272 samples in the validation part and 252 samples in the test part. The balance of positive to negative classes is approximately equal to 1:1;
    
    \item $X_1$ with $P_3$.
    
    It has 22945 samples, with 13247 samples in the train part, 5131 samples in the validation part and 4567 samples in the test part. The balance of positive to negative classes is approximately equal to 1:34, and the majority of the negative class is drawn from the $P_3$ part.    
\end{itemize}

In Appendix~\ref{sec:cwe_distribution}, Figures~\ref{fig:cwe-distribution-with-p3} and \ref{fig:cwe-distribution-without-p3} show the distribution of the top 19 CWE types for the resulting test datasets with and without the $P_3$ part.

\subsection{Evaluation metrics}

We use a standard procedure for model training and evaluation. We train a model for a predefined number of epochs, choose the best checkpoint by ROC AUC on the validation dataset, and evaluate the chosen checkpoint on the test dataset.

The metrics reported include ROC AUC, F1 score for the positive class (vulnerable functions), optimal classification threshold (used for F1 score). The optimal classification threshold is determined on the validation dataset. In most cases, we report both validation and test metrics.

Hyperparameter optimization is performed on the validation dataset using the best validation metrics performance.

\subsection{Pretrained models}

In this work, the primary pretrained model architecture employed is WizardCoder \cite{Luo2023-wizardcoder}, a transformer-based neural network with a decoder-only architecture. With over 13 billion parameters, WizardCoder was pretrained using a causal language modeling objective on a large collection of GitHub source code, endowing the model with extensive knowledge of natural programming language constructs.

We compare WizardCoder to ContraBERT \cite{Liu2023-contrabert}, which is the state-of-the-art model in the CodeBERT family. In contrast to vanilla CodeBERT, ContraBERT introduces an enhanced encoder architecture obtained by cooperation of two pretrained CodeBERT encoders. ContraBERT is pretrained using the usual masked language modeling (MLM) along with a number of contrastive pretraining tasks, and can be finetuned for the defect detection task~\cite{Liu2023-contrabert}. ContraBERT can be considered as an enhancement of the CodeBERT model, which showcases more accurate and robust results than CodeBERT for the defect detection task. Hence, ContraBERT is selected for the experiments.

\subsection{Full reproducibility package}

All necessary code and data for reproducing our experiments are available in our GitHub repository\footnote{Source code and dataset: \url{https://github.com/rmusab/vul-llm-finetune}}.

\section{Research questions}

\subsection{Comparison to CodeBERT-based models.}

\textbf{RQ1: Is a StarCoder-based model more effective than a CodeBERT-based for the balanced vulnerability detection task?}

The models were finetuned on the dataset without easy $P_3$ negatives, using different hyperparameters like the batch size, learning rate, and number of epochs. Optimal settings for this dataset were determined.

\textbf{RQ2: Is a StarCoder-based model more effective than CodeBERT-based for the imbalanced vulnerability detection task?}

Different training recipes, like the number of epochs and loss functions, were compared between the datasets with and without $P_3$ in order to analyze the transferability of hyperparameters.

\subsection{Ablation study.}

\textbf{RQ3: Is the standard LLM training approach effective for vulnerability detection?}

We formulate the vulnerability detection task as a question answering problem where the model predicts the next token in addition to a binary vulnerability label. The goal is to leverage the standard training approach for LLM decoder models, applying it to our classification task. Training utilizes a cross-entropy loss for both the token prediction and vulnerability classification objectives. 

\paragraph{Batch packing properties.} Batch packing enables dynamic batch sizes but may cause unstable gradients. The key questions in this direction are:
\begin{itemize}
\item \textbf{RQ4: Does the dynamic batch size harm the model quality or not?}
\item \textbf{RQ5: Does the mean or the sum loss reduction perform better?}
\end{itemize}

\textbf{RQ6: Does an increased context size provide improvements in quality?}

To determine sensitivity to context, the model was trained with reduced input sequence lengths of 512 tokens instead of 2048 on the dataset without the $P_3$ part. Then, the results were compared in order to analyze impact.

\subsection{Imbalanced classification improvement.}

\textbf{RQ7: Is the focal loss with sample weighting effective for tackling the vulnerability detection class imbalance problem?}

On the full dataset, focal loss and sample weighting techniques were applied to emphasize the minority positive examples from $P_1$. The gamma parameter and sampling weights of the focal loss were systematically varied to ensure the performance.

For all experiments and analyses, the main evaluation metrics were ROC AUC, F1 score, accuracy, and optimal classification threshold. The goal was to make finetuning possible for classification, as well as to determine the best practices for finetuning on an imbalanced task.

\section{Experimental results}

\subsection{Comparison to CodeBERT-based models}

\textbf{RQ1: Is a StarCoder-based model more effective than a CodeBERT-based for the balanced vulnerability detection task?}

We compared the Starcoder-based model with the CodeBERT-based model on the dataset $X_1$ without $P_3$. This dataset is a balanced dataset without easy negative examples. 

The methodology is the following: we take a pretrained model, manually optimize its learning hyperparameters using the training and validation datasets, and then determine the quality of the best model on the test dataset. The optimized hyperparameters include the batch size, learning rate and number of epochs.

The results are presented in Table \ref{tab:rq1_1}. We see that \textbf{the finetuned WizardCoder surpasses finetuned ContaBERT both in ROC AUC and F1 metrics on the balanced vulnerability detection task.} This superior performance of WizardCoder can be attributed to its larger model capacity of 13 billion parameters and pretraining on a much larger code corpus.

\begin{table}[ht]
\centering
\begin{tabular}{lll}
\hline
Base model & ROC AUC & F1 \\
\hline
ContraBERT & 0.66 & 0.68 \\
WizardCoder & 0.69 & 0.71 \\ 
\hline
\end{tabular}
\caption{ROC AUC and F1 scores for finetuned ContraBERT and WizardCoder models on the $X_1$ without $P_3$ dataset.} \label{tab:rq1_1}
\end{table}

To finetune the WizardCoder model, several important hyperparameters were identified: a batch size of $1$ (the actual batch size becomes approximately $120$ by packing functions in the batch), learning rate of $0.0001$, and $50$ epochs of training with cosine annealing schedule. The best checkpoint occurred at epoch $19$, indicating sufficient training time is needed. 

The number of epochs for the cosine schedule is crucial, as it controls the learning rate distribution across epochs and greatly impacts training regimes.

\textbf{RQ2: Is a StarCoder-based model more effective than CodeBERT-based for the imbalanced vulnerability detection task?}

We compared the Starcoder-based model with the CodeBERT-based model on the $X_1$ with $P_3$ dataset. This dataset is an imbalanced dataset with majority of samples belonging to easy negative examples. This dataset is closer to the real-world vulnerability distribution. 

The methodology of comparison is the same as in RQ1. The optimal training hyperparameters appeared to be the same for both datasets ($X_1$ with and without $P_3$). The identical optimal settings imply that the models train similarly on both datasets.

The results are presented in Table \ref{tab:rq2_1}. We see that \textbf{finetuned WizardCoder surpasses finetuned ContaBERT both in ROC AUC and F1 metrics on the imbalanced vulnerability detection task. However, achieved gains over ContraBERT in ROC AUC score are smaller compared to the $X_1$ without $P_3$ dataset}. A potential reason is underutilization of the more informative $P_1$ and $P_2$ partitions on this dataset.

\begin{table}[ht] 
\centering
\begin{tabular}{lll}
\hline
Base model & ROC AUC & F1 \\
\hline
ContraBERT & 0.85 & 0.22 \\
WizardCoder & 0.86 & 0.27 \\
\hline
\end{tabular}
\caption{ROC AUC and F1 scores for finetuned ContraBERT and WizardCoder models on $X_1$ with $P_3$ dataset.} \label{tab:rq2_1}
\end{table}

\subsection{Ablation study}

\textbf{RQ3: Is the standard LLM training approach effective for vulnerability detection?}

The WizardCoder model was finetuned on the vulnerability dataset by formulating the task as a question answering problem. Specifically, the model was trained to predict the next token in addition to the binary vulnerability label on the $X_1$ with $P_3$ dataset. The results are presented in Table \ref{tab:rq3_1}.

\begin{table}[ht]
\centering
\begin{tabular}{lll}
\hline
Base model & Training approach & ROC AUC \\ 
\hline
WizardCoder & Next token prediction & 0.75 \\
CodeBERT & Binary classification & 0.85 \\
WizardCoder & Binary classification & 0.86 \\
\hline
\end{tabular}
\caption{ROC AUC Scores for finetuning with the next token prediction and classification objectives on the imbalanced $X_1$ with $P_3$ dataset.} \label{tab:rq3_1}
\end{table}

The next token prediction approach achieved the ROC AUC score of 0.75 for WizardCoder on the $X_1$ with $P_3$ dataset. \textbf{This result is inferior to CodeBERT and WizardCoder models trained with classification-only objectives (0.85 and 0.86, respectively), but still surpasses random performance.}

\paragraph{Batch packing properties.} The batch packing approach introduces some unique properties:
\begin{itemize}
    \item The actual batch size becomes dynamic during training as sequences are packed;
    \item Using the standard mean loss reduction gives different functions unequal influence on the gradient, which may harm the overall performance;
    \item Using the sum loss reduction training steps unevenly, which may also be harmful to performance.
\end{itemize}

This raises two research questions:
\begin{itemize}
\item \textbf{RQ4: Does the dynamic batch size harm the model quality or not?}
\item \textbf{RQ5: Does the mean or the sum loss reduction perform better?}
\end{itemize}

\textbf{RQ4: Does the dynamic batch size harm the model quality or not?}

The batch packing approach actually introduces the dynamic batch size during training. This dynamic batch size allows to decrease the training time by up to 13 times. Dynamic batch size is not a standard approach for training neural networks, so the following question arises: does it sacrifice the model quality for training speed improvement? In order to answer this question we tried a variant of batch packing that limits the maximum number of functions in a batch. This limiting decreases the dispersion of the actual batch sizes distribution. 

We tested two different values for the maximum number of functions per batch: $50$ and $100$ on the validation part of the $X_1$ without $P_3$ dataset. Our results are present in Table \ref{tab:rq4_1}. The results imply that there is no significant influence of limiting the maximum number of functions per batch, indicating that dynamic batching may not have a detrimental effect on the model’s performance.

\begin{table}[ht]
\centering
\begin{tabular}{ll}
\hline
Max. functions per batch & Validation ROC AUC \\
\hline
No limit & 0.72 \\
 50 & 0.72 \\
100 & 0.72 \\
\hline
\end{tabular}
\caption{WizardCoder's results with limiting the maximum number of functions per batch on the validation part of the $X_1$ without $P_3$ dataset.} \label{tab:rq4_1}
\end{table}

\textbf{In summary, dynamic batch size works well empirically and limiting it does not lead to any improvements.}

\textbf{RQ5: Does the mean or the sum loss reduction perform better?}

Each kind of loss reduction has arguments against it: 
\begin{itemize}
    \item Using the standard mean loss reduction gives different functions unequal influence on the gradient, which may harm the overall performance;
    \item Using the sum loss reduction training steps unevenly, which may also be harmful to performance.
\end{itemize}

Thus, it becomes important to investigate whether the mean or sum loss reduction is better suited for the batch packing approach. In order to answer this question, we performed testing on the validation part of the $X_1$ without $P_3$ dataset. The results are present in Table \ref{tab:rq5_1}. The mean loss reduction performed poorly, highlighting issues with the mean loss in comparison with the sum reduction. 

The superior performance of the sum loss reduction indicates that it better handles the uneven sequence lengths arising from the batch packing. By summing over the packed sequences, each function contributes equally to the gradient.

\begin{table}[ht] 
\centering
\begin{tabular}{ll}
\hline
Loss reduction & Validation ROC AUC \\
\hline
Mean  & 0.57 \\
Sum  & 0.72 \\
\hline
\end{tabular}
\caption{WizardCoder's performance by a loss reduction method on the validation part of the $X_1$ without $P_3$ dataset.} \label{tab:rq5_1}
\end{table}

\textbf{In summary, the sum loss reduction method outperforms the mean one by better handling uneven sequence lengths.}

\textbf{RQ6: Does an increased context size provide improvements in quality?}

An open question was whether the quality improvements obtained by WizardCoder stemmed from the larger 2048 context size versus the 512 baseline, or better code understanding by the model.

We conducted an experiment to isolate the impact of context size. The 2048 context size WizardCoder model was compared to the 512 context size version on the test part of the $X_1$ without $P_3$ dataset.

\begin{table}[ht]
\centering
\begin{tabular}{ll}
\hline
Context size & Test ROC AUC \\
\hline
512 Tokens & 0.69 \\
2048 Tokens & 0.69 \\ 
\hline
\end{tabular}
\caption{The impact of the WizardCoder's context size on performance.} \label{tab:rq6_1}
\end{table}

Table \ref{tab:rq6_1} shows that \textbf{reducing the context size from 2048 tokens to 512 tokens resulted in the identical validation ROC AUC score of 0.69 for WizardCoder. This suggests the performance gains are mainly due to the model's learning of improved code representations, rather than the increased context size.}

\subsection{Imbalanced classification improvement}

\textbf{RQ7: Is the focal loss with sample weighting effective for tackling the vulnerability detection class imbalance problem?}

The obtained gains of the StarCoder-based model over the CodeBERT-based on the imbalanced $X_1$ with $P_3$ dataset are minor compared to the $X_1$ without $P_3$ dataset. A potential reason is underutilization of the more informative $P_1$ and $P_2$ parts on this dataset. In order to explore the influence of better utilization of $P_1$ and $P_2$ parts on the model's quality, we conducted experiments that incorporate the focal loss and sample weighting techniques:

\begin{itemize}
    \item Focal loss emphasizes hard and rare examples contained in the $P_1$ and $P_2$ parts;
    \item Sample weighting also focuses model on the minority cases from the $P_1$ and $P_2$ parts.
\end{itemize}

\paragraph{Focal Loss.}
Focal loss \cite{Lin2017-focalloss} applies a modulating factor $(1 - p_t)^\gamma$ to the standard cross entropy loss. This factor down-weights well-classified examples and focuses training on hard misclassified cases. The standard cross-entropy loss corresponds to $\gamma=0$.

We tested $\gamma$ values from $1$ to $5$ on the validation part of the $X_1$ with $P_3$ dataset. The results are presented in Table \ref{tab:rq7_1}.

\begin{table}[ht]
\centering
\begin{tabular}{lcccc}
\hline
& $\gamma=0$ & $\gamma=1$ & $\gamma=3$ & $\gamma=5$ \\
\hline
ROC AUC & 0.858 & 0.873 & 0.868 & 0.852 \\
F1 & 0.277 & 0.265 & 0.272 & 0.250 \\
Best val. epoch & 12 & 12 & 17 & 11 \\
Best threshold & 0.087 & 0.288 & 0.265 & 0.4 \\
\hline
\end{tabular}
\caption{Resulting quality of WizardCoder for different $\gamma$ values in focal loss on the imbalanced $X_1$ with $P_3$ dataset.} \label{tab:rq7_1}
\end{table}

In summary, the focal loss with $\gamma=1$ achieves a small improvement in ROC AUC over the standard cross entropy loss ($\gamma=0$). However, the gains are small.

The peak quality at $\gamma=1$ followed by a decrease at higher $\gamma$ values suggests the importance of balancing the emphasis on hard examples with retaining sufficient training signal from easier instances.

One potential reason for such small gains is that the noisy or imperfect labels limit the ability of the focal loss to accurately identify the most important hard examples to prioritize. If the label of an example does not precisely reflect its difficulty, a heavy focus on hard cases may be less effective.

The inconsistent improvements demonstrate the need for further exploration of methods to leverage hard examples without detriment to learning on easier cases, while taking the label quality into account.

An interesting observation is that the focal loss leads to better calibrated models, with thresholds closer to $0.5$. This indicates that the predicted class probabilities are closer to the true values under the focal loss. The better calibration is a useful characteristic.

\paragraph{Adding sample weights.}

Sample weighting assigns higher importance weights to the $P_1$ and $P_2$ examples compared to $P_3$ during training. Weights from $1.0$ to $30.0$ were tried. Larger weights place more emphasis on the minority informative data. The similar technique was reported to be effective with the focal loss in the original article \cite{Lin2017-focalloss}.

The applied technique is different from the usual class weighting scheme since it puts more weight on both the samples of the positive ($P_1$) and negative ($P_2$) parts. In the previous experiments, we found that the class weighting technique was ineffective, so the sample weighting scheme provides a different perspective.

The numeric results are presented in Table \ref{tab:rq7_2}.

\begin{table}[ht]
\centering
\begin{tabular}{llllll}
\hline
Weight & \thead{Best val.\\ AUC} & \thead{Best \\ epoch} & \thead{Test \\ AUC} & \thead{Test \\ F1} & Threshold \\
\hline
30   & 0.876 & 6 & 0.863 & 0.22 & 0.42 \\
10.0 & 0.878 & 13 & 0.86 & 0.24 & 0.44 \\
3.0  & 0.88 & 12 & 0.875 & 0.265 & 0.07 \\
1.0 (base) & 0.87 & 12 & 0.858 & 0.277 & 0.08 \\
\hline
\end{tabular}
\caption{Resulting quality of WizardCoder for different values of $P_1 + P_2$ weights on the imbalanced $X_1$ with $P_3$ dataset.}
\label{tab:rq7_2}
\end{table}

From Table \ref{tab:rq7_2} we conclude that adding sample weights for the $P_1$ and $P_2$ partitions can provide small improvements in ROC AUC and F1 score over the baseline. However, large weight values degrade performance.

The best weighting scheme of $3x$ attains marginal gains in both ROC AUC and F1 over unweighted training. However, excessive weighting, like $30x$, hurts the validation and test metrics.

This indicates there is an optimal moderate weighting that slightly emphasizes the hard and rare $P_1$ and $P_2$ examples without skewing the distribution too heavily. Further tuning of the weighting factor may yield additional gains.

\paragraph{Combination of sample weights with focal loss}

Both the focal loss and additional weights do similar work. In the original work \cite{Lin2017-focalloss}, it was stated that adding weights to the focal loss leads to some improvements.

Therefore, we combined the focal loss with the sample weighting technique. The results are presented in Table \ref{tab:rq7_3}. 

\begin{table}[ht]
\centering
\begin{tabular}{lllll}
\hline
Weight & $\gamma$ & Val. AUC & Test AUC & Test F1 \\
\hline
3 & 1.0 & 0.88 & 0.877 & 0.273 \\
10 & 3 & 0.872 & 0.853 & 0.226 \\
10 & 1 & 0.877 & 0.863 & 0.243 \\
3 & 3 & 0.874 & 0.865 & 0.286 \\
None & 1 & 0.87 & 0.873 & 0.265 \\
None & 3 & 0.87 & 0.868 & 0.272 \\
\hline
\end{tabular}
\caption{The results of WizardCoder using the focal loss and the sample weighting on the imbalanced $X_1$ with $P_3$ dataset.} \label{tab:rq7_3}
\end{table}

\textbf{Experiments with the focal loss and sample weighting demonstrated minor improvements over the baseline training, with the best model achieving 0.877 ROC AUC versus 0.86 for the baseline. However, neither technique provided substantial gains.} More advanced methods that can explicitly account for a severe class imbalance are needed.

\section{Threats to validity}

There are some potential threats to validity that could limit the objectiveness of our study:

\begin{itemize}
    \item The dataset is not split by projects, and splitting by projects might have resulted in a significant decrease in quality;
    \item When finetuning a model in the classification regime with batch packing, the model is able to see the functions that precede the current one in the input. This could limit the training effectiveness of the resulting model. However, the negative impact on training should be mitigated since these functions are placed randomly, so it would be hard for the model to learn irrelevant features. Also, during inference for test data, only one is function was considered in any single input sequence, so there is no bias in the test scores;
    \item The function-level granularity of our dataset limits the types of vulnerabilities that span across multiple functions;
    \item The dataset size is relatively small. This might limit its representativeness of the true vulnerability distribution.
\end{itemize}

\section{Conclusion}

Our work demonstrates the effectiveness of finetuning large language models for the vulnerability detection problem in source code. The WizardCoder model achieved state-of-the-art results, with the ROC AUC score of $0.69$ on the dataset without easy negative examples. This improves over previous CodeBERT-based models, likely due to the WizardCoder's larger model capacity and pretraining corpus.

Several key contributions are made:

\begin{itemize}
\item An efficient batch packing strategy is developed to mitigate small sequence lengths, providing over $13x$ speedup in training time. This enables faster iteration and tuning;

\item An improvement in ROC AUC from $0.66$ to $0.69$ was obtained over the state-of-the-art non-LLM model ContraBERT on the $X_1$ dataset without $P_3$ part. An improvement of the F1 score was obtained for the dataset with $P_3$ ($0.27$ vs $0.22$);


\item For the highly imbalanced dataset, it was shown that the focal loss with sample weighting gives improvements from $0.86$ to $0.878$. Despite these improvements, more advanced methods are needed to properly emphasize the minority positive examples;

\item A new more precise vulnerability benchmark dataset is collected. It is smaller than most of the available datasets, as well as possesses quality labels that are free from the errors stemming from irrelevant code and cleanup changes \cite{Croft2023-dataquality}.
\end{itemize}

Opportunities remain for further improvements in this approach, mainly related to training on the dataset with the $P_3$ part included. Future work should explore techniques like curriculum learning, active sampling, and data augmentation to better leverage the scarce minority data. The insights gained can guide research on broader tasks involving code analysis and understanding.

We show an improvement in quality of the vulnerability detection task by finetuning WizardCoder LLM model for the vulnerability detection task. Considering the quality needed for business tasks, there is still a gap between the achieved and required quality. This leaves us with a big room for improvement, particularly, in choosing the optimal representation of vulnerabilities and usage of a broader project-level context information. These are very exciting prospects for future research.


\bibliography{main}
\bibliographystyle{icomp2024_conference}

\begin{appendices}

\section{Distribution of CWEs} \label{sec:cwe_distribution}

\begin{figure}[h]
\centering
\begin{tikzpicture}
\begin{axis}[
    width=0.9\textwidth,
    height=0.4\textheight,
    ybar,
    bar width=20pt,
    enlargelimits=0.01,
    legend style={at={(0.5,-0.15)}, anchor=north, legend columns=-1},
    ylabel={Frequency},
    symbolic x coords={
        Other,CWE-862,CWE-200,CWE-863,CWE-352,CWE-89,CWE-611,CWE-79,CWE-22,
        CWE-776,CWE-264,CWE-20,CWE-835,CWE-295,CWE-287,CWE-203,CWE-755,
        CWE-444,CWE-119
    },
    xtick=data,
    x tick label style={rotate=45,anchor=east},
    nodes near coords,
    nodes near coords align={vertical},
    every node near coord/.append style={font=\normalsize},
]
\addplot coordinates {
    (Other,902) (CWE-862,504) (CWE-200,429) (CWE-863,391) (CWE-352,324)
    (CWE-89,273) (CWE-611,261) (CWE-79,201) (CWE-22,194) (CWE-776,175)
    (CWE-264,153) (CWE-20,142) (CWE-835,125) (CWE-295,91) (CWE-287,90)
    (CWE-203,85) (CWE-755,81) (CWE-444,81) (CWE-119,74)
};
\end{axis}
\end{tikzpicture}
\caption{Distribution of the top 19 CWE categories and others in the test set with $P_3$.}
\label{fig:cwe-distribution-with-p3}
\end{figure}

\begin{figure}[h]
\centering
\begin{tikzpicture}
\begin{axis}[
    width=0.9\textwidth,
    height=0.4\textheight,
    ybar,
    bar width=20pt,
    enlargelimits=0.01,
    legend style={at={(0.5,-0.15)}, anchor=north, legend columns=-1},
    ylabel={Frequency},
    symbolic x coords={
        Other,CWE-79,CWE-611,CWE-20,CWE-22,CWE-264,CWE-502,CWE-200,CWE-94,
        CWE-863,CWE-287,CWE-74,CWE-78,CWE-284,CWE-276,CWE-755,CWE-89,
        CWE-269,CWE-352
    },
    xtick=data,
    x tick label style={rotate=45,anchor=east},
    nodes near coords,
    nodes near coords align={vertical},
    every node near coord/.append style={font=\normalsize},
]
\addplot coordinates {
    (Other,71) (CWE-79,36) (CWE-611,23) (CWE-20,22) (CWE-22,13)
    (CWE-264,10) (CWE-502,9) (CWE-200,8) (CWE-94,8) (CWE-863,8)
    (CWE-287,6) (CWE-74,6) (CWE-78,6) (CWE-284,6) (CWE-276,4)
    (CWE-755,4) (CWE-89,4) (CWE-269,4) (CWE-352,4)
};
\end{axis}
\end{tikzpicture}
\caption{Distribution of the top 19 CWE categories and others in the test set without $P_3$.}
\label{fig:cwe-distribution-without-p3}
\end{figure}
\clearpage

\end{appendices}

\end{document}